\documentclass[11pt]{article}

\usepackage{graphicx}
\usepackage{color, colortbl}
\usepackage[caption=false,font=footnotesize]{subfig}
\usepackage{amsmath, amssymb}
\usepackage{enumitem}
\usepackage{multirow}

\usepackage{authblk}

\begin{document}
	
\title{Market Graph Clustering Via QUBO and Digital Annealing}

\author[1]{Seo Hong}
\author[2]{Pierre Miasnikof \thanks{corresponding author: p.miasnikof@mail.utoronto.ca}}
\author[1]{Roy Kwon}
\author[2]{Yuri Lawryshyn}
\affil[1]{Dept. of Mechanical and Industrial Engineering, University of Toronto, Toronto, ON, Canada}
\affil[2]{Dept. of Chemical Engineering and Applied Chemistry, University of Toronto, Toronto, ON, Canada}

\date{}

\maketitle

\begin{abstract}
Our goal is to find representative nodes of a market graph that best replicate the returns of a broader market graph (index), a common task in the financial industry. We model our reference index as a market graph and express the index tracking problem in a quadratic K-medoids form. We take advantage of a purpose built hardware architecture, the Fujitsu Digital Annealer, to circumvent the NP-hard nature of the problem and solve our formulation efficiently. In this article, we combine three separate areas of the literature, market graph models, K-medoid clustering and quadratic binary optimization modeling, to formulate the index-tracking problem as a quadratic K-medoid graph-clustering problem. Our initial results  show we accurately replicate the returns of a broad market index, using only a small subset of its constituent assets. Moreover, our quadratic formulation allows us to take advantage of recent hardware advances, to overcome the NP-hard nature of the problem.
\end{abstract}


\section{Introduction}
Our work is an empirical implementation of the K-medoid clustering technique expressed as a quadratic unconstrained binary optimization model (QUBO) and applied to market graphs. It is inspired by the seminal work of Boginski et al \cite{BogStruct2003,BogStat2005}, Cornu\'{e}jols et al. \cite{CornOptFin2007} and that of Bauckhage et al. \cite{Bauck2019}. We combine these pieces of complementary but disjoint work, to formulate the index-tracking problem as a QUBO K-medoid clustering of a broader market graph problem.

Graph clustering is an unsupervised learning task, consisting of assigning common labels to vertices deemed similar. It has found applications in many areas. Chemistry, biology, social-networks and finance are a few examples where graph clustering has been applied. However, while there are many competing techniques, the graph clustering problem is NP-hard, which limits its scope of application.   

QUBO formulations of many mathematical problems have recently gained in interest. This recent spike in interest is, in no small part, due to recent advances in computer hardware and the availability of purpose-built hardware for their solution that circumvent the NP-hard nature of the problem. Examples of this novel hardware are Fujitsu's Digital Annealer (DA) and D-Wave's Quantum Annealer.

Graphs have recently been introduced as models of the stock market. In addition, clustering of stock market data is a longstanding focus of interest for both practitioners and academics. It has been used for various purposes, like risk management and portfolio diversification, for example. Index-tracking is another longstanding interest in finance. It consists of building tracking-portfolios whose returns follow a broader index's return, but with a subset of stocks. Some authors in the field have used clustering for the purpose of index-tracking. Their methods identify exemplars of subsets of an index and construct tracking-portfolios consisting of only those exemplars.

Our initial results are very encouraging. Our tests show we accurately replicate the returns of a broad market index, using only a small subset of its constituent assets. Moreover, our QUBO formulation allows us to take advantage of recent hardware advances, to overcome the NP-hard nature of the problem.

\section{Previous Work}
Our work lies at the intersection of graph models of the stock markets,  clustering, combinatorial optimization (QUBO) and index tracking. In this section, we briefly review these four areas of research. Our goal is not to provide the reader with a detailed review of the state of the art in these very broad fields, but rather to focus specifically on their relevance to the work in this article, in order to put it in context.

The use of graphs as models of the stock market is initially introduced in the literature by the very extensive work of Boginski et al. \cite{BogStruct2003,BogNetMassive2004,BogNetStock2004,BogStat2005,BogNetMin2006}. While different methods have been suggested for determining edge-weights, the idea is to model stocks as vertices and assign edge weights proportional to their returns correlations. 

Other authors have also followed up on and expanded this work by studying graph dynamics over time  \cite{ArratiaTempor2011,KochetDynamics2014} and examined methods for building the graph \cite{BautinSimil2013,KoldStat2013,KalyaginOptDec2018}. In fact, to this day, the topic of graphs as a model for equity markets remains a subject of discussion in the literature \cite{AbramsGraphEquity2016,MartiReview2019}.

Graph clustering is the process of assigning common labels to vertices deemed similar. It has a long history in the literature. A thorough review of the graph clustering literature is beyond the scope of this article. For a very comprehensive view of the field, we refer the reader to
the foundational work of Schaeffer \cite{Schaeffer2007}, Fortunato \cite{FortunatoLong2010} and the recent contribution by Fortunato and Hric \cite{guideFortunato16}.

The link between clustering and portfolio construction is of particular relevance to the work in our article \cite{CornOptFin2007,Chen2012,BogNet2014,KalyaginMarkow2014,Wu2017,PuertoPortSelect2020}. Although not focused specifically on graph clustering, Cornu\'{e}jols et al. present a K-medoid formulation for index-tracking \cite{CornOptFin2007}. These authors use the standard K-medoid technique \cite{ESL2009} to find `$K$' representative stocks that compose a portfolio that replicates a broader index. 

More recently, Bauckhage et al. \cite{Bauck2019} reformulate the K-medoids problem in QUBO form \cite{IsingForm2014,Glover2018}. This reformulation allows us to take advantage of novel purpose-built hardware specifically designed for QUBO formulations \cite{physInspired2019,DAU2020}.

\section{Methods}
We begin with a market graph consisting of $n=453$ stocks that have been constituents of the Standard and Poors 500 index (SP500) for every year since 2014. We apply a K-medoid index-tracking technique to find `$k=10$' exemplars that will form our tracking portfolio. Finally, to take advantage of fast purpose-built computer hardware, the Fujitsu DA, we express the K-medoid problem as a QUBO.

\subsection{Market Graph}
We represent our universe of $n=453$ stocks as a complete weighted graph, where edge weights represent the association between stocks `$i$' and `$j$'. These weights are defined using the Pearson correlation coefficient of the log daily returns,
\[d_{ij} = \sqrt{2 \left(1-\rho_{ij} \right)} \; .\]
For each year in our study, edge weights are recomputed annually, using the previous year's daily returns. In modeling the market in this way, our investment universe is modeled as a complete weighted graph, with no self-loops (since $\rho_{ii} = 1$).

To be consistent with the QUBO formulation of Bauckhage et al. \cite{Bauck2019}, we convert our adjacency (distance) matrix into a more robust matrix $\Delta=[\delta_{ij}]$, with the elements $\delta_{ij} = 1 - \exp(-\frac{1}{2} \times d_{ij})$. We note that this formulation requires all-pairs distances ($d_{ij}$) be known, which is why we use a complete graph representation.

\subsection{QUBO Model} \label{modelsection}
Putting it all together, we formulate our K-medoid problem of finding a portfolio of $k=10$ exemplars to replicate the returns of the $n=453$ constituents of the SP500 as 
\begin{equation} \label{model} 
\begin{aligned}
& \underset{\vec{z}}{min} \quad \vec{z}^T\left(\gamma \mathbf{1}\mathbf{1}^T - \alpha \frac{1}{2}\Delta\right)\vec{z} + \vec{z}^T\left(\beta \Delta\mathbf{1} - 2\gamma k\mathbf{1}\right)\\
& z_i \in \{0,1\}, \quad \forall i \in V \\ 
&\text{where, } \mathbf{1} \text{ denotes a vector of ones of appropriate dimension.}
\end{aligned}
\end{equation}
Our model \eqref{model} consists $n$ decision variables, $z_i=1$ if node $i$ is an exemplar node and 0 otherwise. We follow the example of Bauckhage et al. \cite{Bauck2019} and set the parameters $\alpha=\frac{1}{k}$, $\beta=\frac{1}{n}$, $\gamma=2$. 
(For more details on this QUBO formulation, we refer the reader to the original work of Bauckhage et al. \cite{Bauck2019}.)

\subsection{The Fujitsu DA: Purpose-Built Hardware}
To circumvent the NP-hard nature of the clustering problem, we use purpose built architecture, the Fujitsu DA. The DA provides fast computation and is designed specifically for combinatorial optimization problems expressed in QUBO form \cite{physInspired2019,DAU2020}.

All our computations for the minimization of the model described in Section~\ref{modelsection} were done using this architecture. More specifically, these computations were done using hardware built exclusively for the University of Toronto's research environment.

\section{Numerical Experiments}
We use our K-medoid technique to construct four index-tracking portfolios, one for each year in our sample (2016-19). For each year, we use the previous year's returns to compute stock-to-stock distances, build a new market graph and corresponding matrix $\Delta$. We then optimize the QUBO model, using the DA, to obtain a tracking portfolio.

To assess tracking accuracy, we use tracking-error and ``beta to the index'', both measured with respect to the full SP500 in each year, as per industry practice.  For each year, we compute the differences between the daily log returns of the SP500 benchmark and of the tracking portfolio. We calculate the standard deviation of the differences to obtain the annual tracking-error. We also regress index returns on market returns to obtain the ``beta'' of the tracking portfolio, the slope of the regression line.

\subsection{Performance Measure: Tracking-Error}
Tracking-error is the standard deviation of the differences between each pair of observations at a given time point (daily in this case). We denote it as $\epsilon$ and compute it as 
\begin{eqnarray*}
d &=& r_{\text{index}} - r_{\text{port}} \\
\epsilon &=& \sqrt{Var[d]} \; .
\end{eqnarray*}

\subsection{Performance Measure: ``beta''}
The ``beta'' of the portfolio is the slope coefficient of the regression of its returns on market returns. A portfolio that tracks the index perfectly has a ``beta'' of one. The regression model we fit to obtain the ``beta'' is 
\[
r_{\text{port}} = \alpha + \beta \times r_{\text{index}} \; .
\]

\subsection{Empirical Results}
\begin{figure}[h!]
	\centering
	\subfloat[2016]{ \includegraphics[width = 0.5\textwidth]{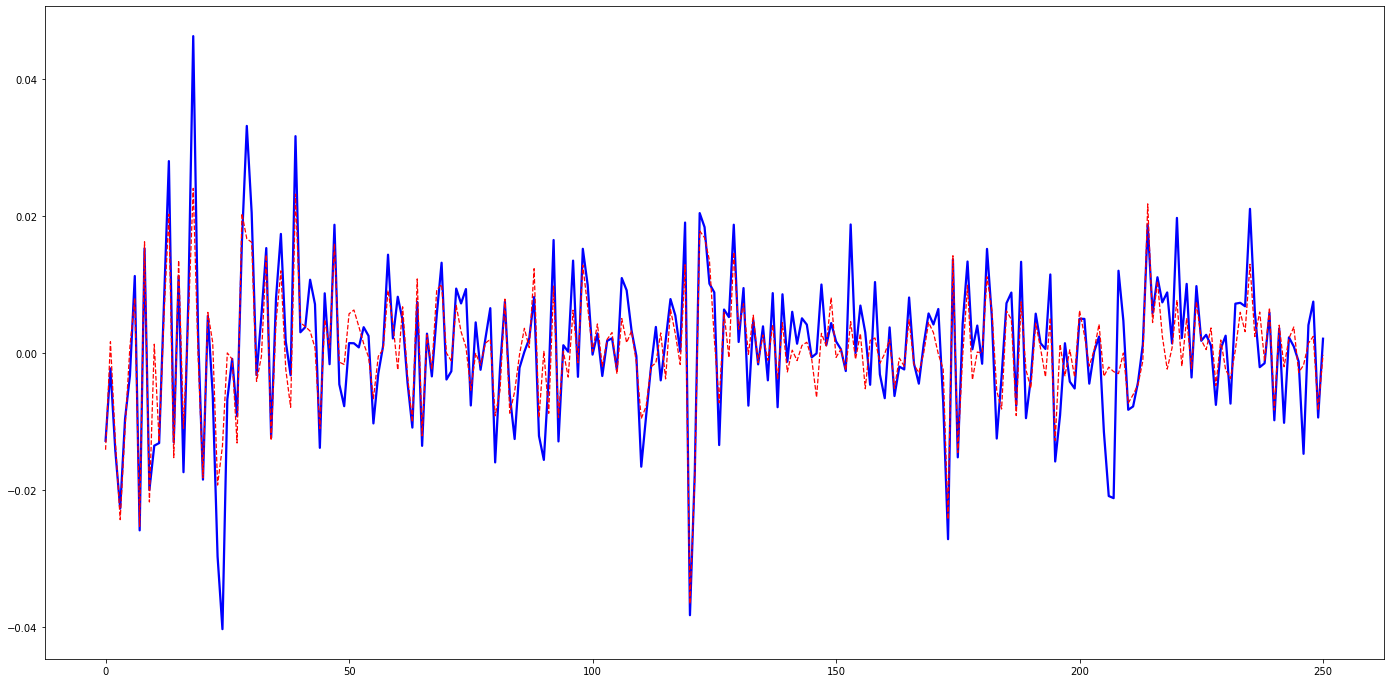} }
	\subfloat[2017]{ \includegraphics[width = 0.5\textwidth]{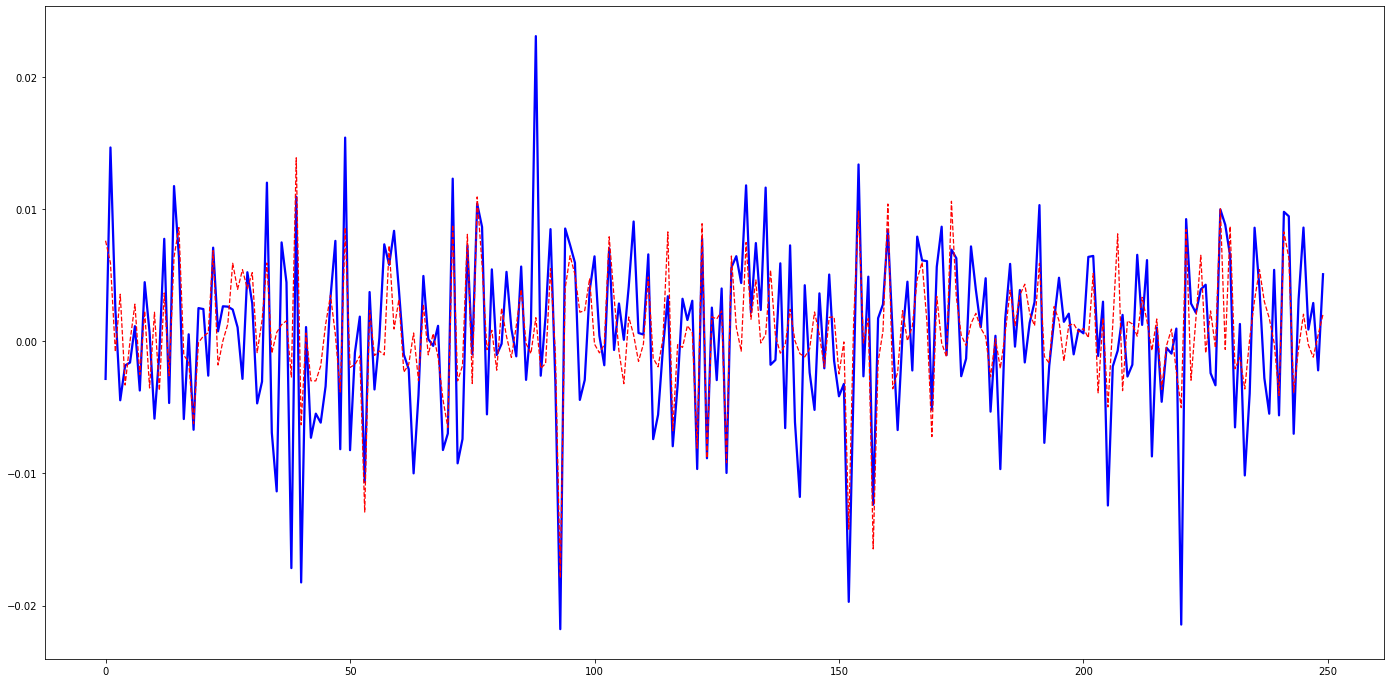} } 
\\
	\subfloat[2018]{ \includegraphics[width = 0.5\textwidth]{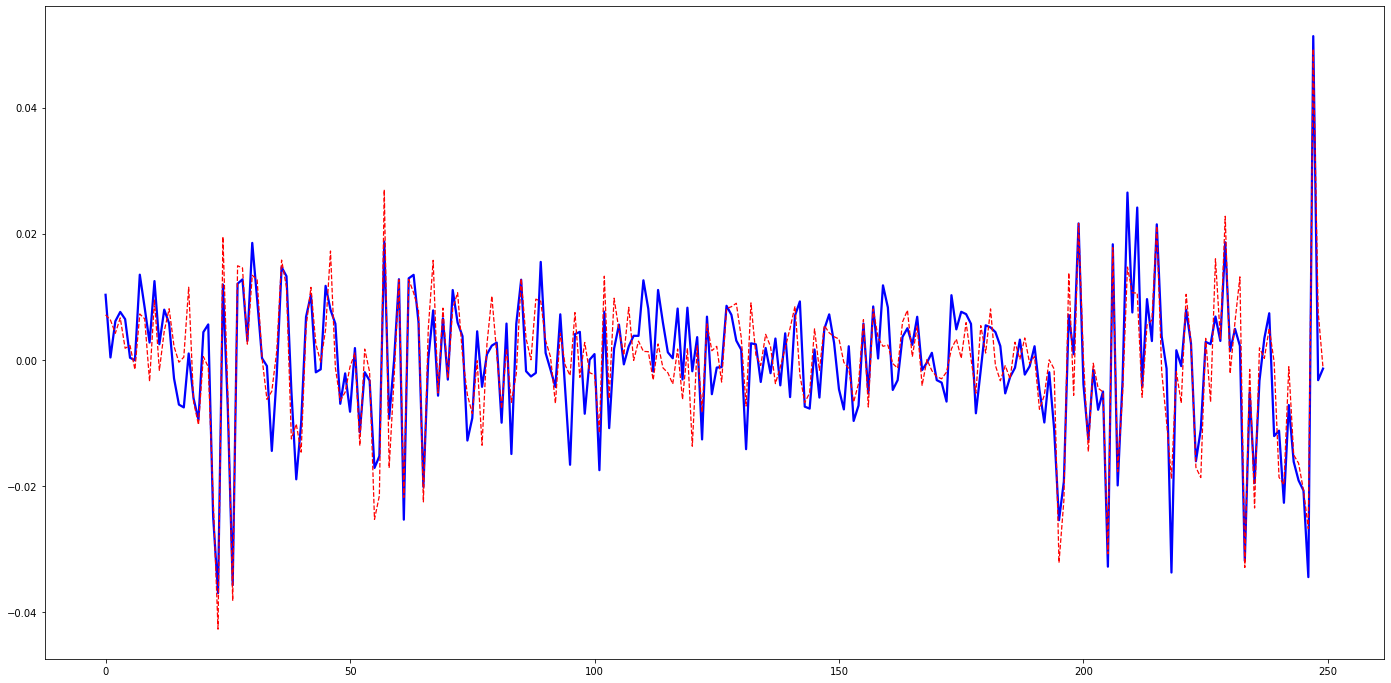} }
	\subfloat[2019]{ \includegraphics[width = 0.5\textwidth]{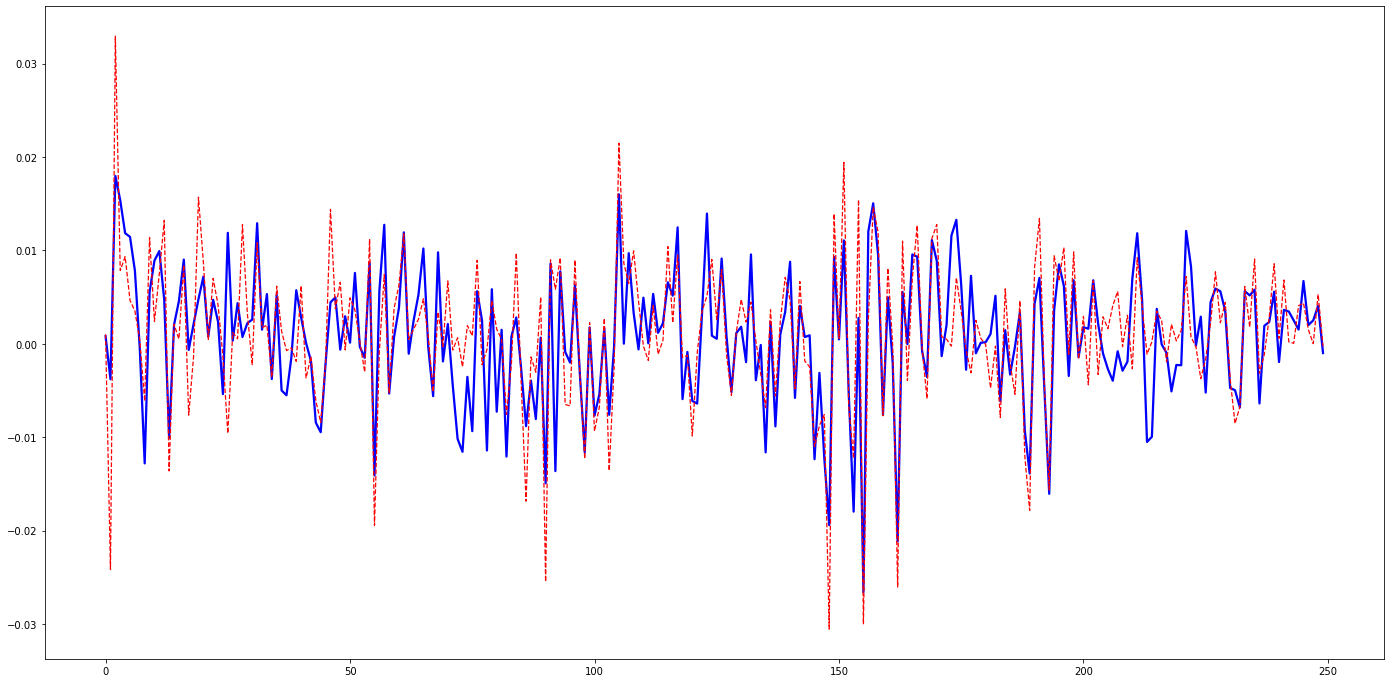} } 
	\caption{Daily Log-Return Series for the Tracking Portfolio and SP500}
	\label{series}
\end{figure}

Tracking error, the ``beta'' and associated t-statistic are reported for each year, in Table~\ref{res}.
\begin{table}[h!]
	\caption{Annual tracking performance}\label{res}
	\centering
	\begin{tabular}{cccc}
		\hline
		\hline
		Year &  Tracking Error & Beta &  t-stat \\
		\hline
		2016 & 0.005 & 1.167 & 29.271 \\
		2017 & 0.005 & 1.068 & 15.482 \\
		2018 & 0.005 & 0.902 & 29.996 \\
		2019 & 0.005 & 0.691 & 18.990 \\
		\hline
		\hline
	\end{tabular}\\[5em]
\end{table}
Daily log-returns for the replicating portfolio (solid blue line) and SP500 (dotted red line) are shown in Figure~\ref{series}.
 
\section{Conclusion and Future Work}
Our results show that a QUBO formulation of the K-medoid problem can be successfully used to replicate a broad market index, using just a few assets. Using only a subset of ten assets, we are able to track the SP500 daily returns with a tracking error of less than 1\%.

On the empirical side, future work will focus on alternate techniques for building the market graph and determining the optimal cardinality of the tracking subset. From a mathematical and computational point of view, we also intend to investigate alternate problem formulations and larger scale optimization.

\section*{Acknowledgements}
We would like to thank Fujitsu Laboratories Ltd and Fujitsu Consulting (Canada) Inc for providing financial support and access to the Digital Annealer at the University of Toronto.

\bibliography{myBib2}

\end{document}